\documentclass[
  preprint,         
  superscriptaddress, 
  aps,             
  prl,             
  nobibnotes,      
]{revtex4-2}

\usepackage{graphicx}
\usepackage{bm}
\usepackage{hyperref}
\usepackage{amsmath}
\usepackage{amssymb}
\usepackage{makecell}
\usepackage{rotating}
\usepackage{tabularray}
\usepackage{xcolor}
\usepackage{natbib}
\usepackage{float}

\begin{document}

\title{Pulsatile poromechanics in layered soft media controls fluid flow and solute transport: from fundamentals to brain clearance}

\author{Matilde Fiori}

    \affiliation{Institut de Mécanique des Fluides de Toulouse, IMFT,
Université de Toulouse, CNRS, Toulouse, 31400, France }

\author{Sylvie Lorthois}
  \affiliation{Institut de Mécanique des Fluides de Toulouse, IMFT,
Université de Toulouse, CNRS, Toulouse, 31400, France }

\date{\today}

\begin{abstract}
Soft porous media often feature a heterogeneous structure. Notably, biological tissues --- such as cartilage and the brain tissue --- consist of two or more layers, with varying mechanical and fluid-flow properties. 
Despite the ubiquity of periodic loading in these systems, the physical implications of layering on nonlinear poromechanics and solute transport remain poorly understood. Uncovering these coupled mechanisms could clarify the fundamental physics behind pressing topics, such as brain metabolic clearance.
Here, we address this gap using a bilayer model of a generic soft porous medium. To isolate the specific role of layering, we select combinations of material properties (porosity, permeability, and \textit{p}-wave modulus) that maintain an identical poroelastic timescale, $T_{\mathrm{PE}}$, across four layered configurations and a reference homogeneous case. We demonstrate that while $T_\mathrm{PE}$ is the key parameter governing the response in a homogeneous medium, the same $T_\mathrm{PE}$ leads to non-trivial localization/propagation patterns for strain, fluid flow and solute transport in a bilayer medium. These non-intuitive results suggest that layered architectures may provide functional benefits for cellular homeostasis over homogeneous ones. Finally, we show that pathological alterations to the brain’s layered structure significantly disrupt fluid-flow and metabolic waste clearance, offering a possible mechanical explanation for impaired transport in disease.

\end{abstract}

\maketitle


\textcolor{green}{}
\section{Introduction}
Numerous soft porous media feature a layered structure. In biological tissues, layered architectures can arise from collagen fibre alignment, which provide increased strength and elasticity along preferred load-bearing directions~\citep{fung2013biomechanics}, or from differences in material composition. The brain, for example, is divided into grey and white matter at the macro-scale, while, at the meso-scale, perivascular spaces --- regions of increased porosity surrounding blood vessels --- form high permeability pathways, facilitating cerebrospinal fluid flow and thus playing an essential role in brain clearance from metabolic waste~\citep{Ilif2012, Rasmussen22}. 

Such division in layers provides space-dependent mechanical and fluid-flow properties, such as solid matrix stiffness, porosity and permeability. This is well studied in several different tissues, from articular cartilage \citep{Park2003, Krishnan2003} to the brain at the macro~\cite{BUDDAY2015} and meso-scale \cite{kedarasetti-fbcns-2022, Causemann2025}.
In cartilage, these layers directly influence solute transport \citep{Zhang2007}, as depth-dependent properties enhance advective and diffusive fluxes at the local layer scale~\citep{Zhang2011, Arbabi2015, DiDomenico2017,Ateshian2010}.
Furthermore, layers in soft tissues may provide functional advantages at the whole tissue scale~\cite{Hashemi2021}: in tissue-engineering, heterogeneous scaffolds with layered or permeability-graded architectures promote cell growth compared to homogeneous designs~\cite{Camarero-Espinosa2016, Mesallati2013, Sano2018}.
Conversely, disrupting organisation can have deleterious consequences: in osteoarthritis, damage to superficial cartilage layers can trigger feedback mechanisms that exacerbate global tissue degeneration~\citep{Hwang2008,Stender2017}.
Guided by the analogy with layer-dependent bio-poromechanics in cartilage, we investigate the controversial biophysics of brain clearance: we hypothesise that layers in the brain could play a role in enhancing fluid-flow and waste transport, and that pathological modifications may disrupt them.   
Abnormal accumulations of metabolic waste, such as amyloid-$\beta$ associated with Alzheimer's disease and other neuropathologies, indeed lead to alterations of the grey matter properties \citep{HRABETOVA2004, Sykova2005, Murphy2011, MURPHY2016, Hiscox2019}, which may disrupt the poromechanical coupling with the adjacent layer (white matter at the macro-scale and/or perivascular spaces at the meso-scale). As in damaged cartilage, this may alter the overall deformation driven by blood vessel pulsations, well known as regulators of brain clearance \cite{VanVeluw2020, Mestre18} and trigger further toxic waste accumulation.

Understanding these complex physiological systems requires extending the framework of nonlinear poromechanics to account for macroscopic heterogeneities (layers). Rather than aiming for a homogenized multilayer medium \cite{Rajapakse95, Kudarova2013, Liu2013}, our objective is to capture the specific effects of macroscopic heterogeneity itself.
While the response of homogeneous media to static~\cite{Macminn2016, hewitt2016, Hewitt2016pre} and periodic loading~\cite{Fiori2023, Cacheux23} --- and the resulting solute transport~\cite{Fiori2025} --- is well-characterised, the fundamental role of macroscopic heterogeneities remains largely unexplored. Periodic loading is particularly relevant here, not only for its biological ubiquity but also for the unique space-dependent \textit{squishy} dynamics it triggers in soft media~\cite{Worster_2024}.
Moreover, the spatial cyclic poromechanical response can be significantly altered by local degradation~\cite{Godard25} or reinforcement~\cite{Alcaide2026Arxiv} of material properties. Building on this, we propose that macroscopic heterogeneities, such as layering, fundamentally reshape the distribution of stress, strain, fluid flow, and the resultant solute transport under large-deformation oscillations. Elucidating these effects is critical to understanding the homeostatic equilibrium of layered biological tissues.

This study is organized as follows. We begin by isolating the structural influence of layering on the overall poromechanical response under periodic loading conditions.
To this end, we adopt a simplified bilayer configuration of equal thickness. By fixing the poroelastic timescale across all setups, we systematically evaluate four distinct layered cases against a baseline homogeneous medium.
Building upon these mechanical insights, we then evaluate how these architectural variations modulate solute transport dynamics. Finally, we demonstrate the practical utility of our findings by applying the developed framework to a realistic physiological system: the cerebral grey matter–white matter bilayer. Evaluating this specific system provides essential insights into the physical mechanisms governing brain clearance, offering a foundation for understanding how they are disrupted in pathological conditions.

\section{Isolating the role of layering: definition of the case studies}

\begin{figure}
    \centering
    \includegraphics[width=0.25\textwidth, trim=3cm 0cm 3cm 0cm]{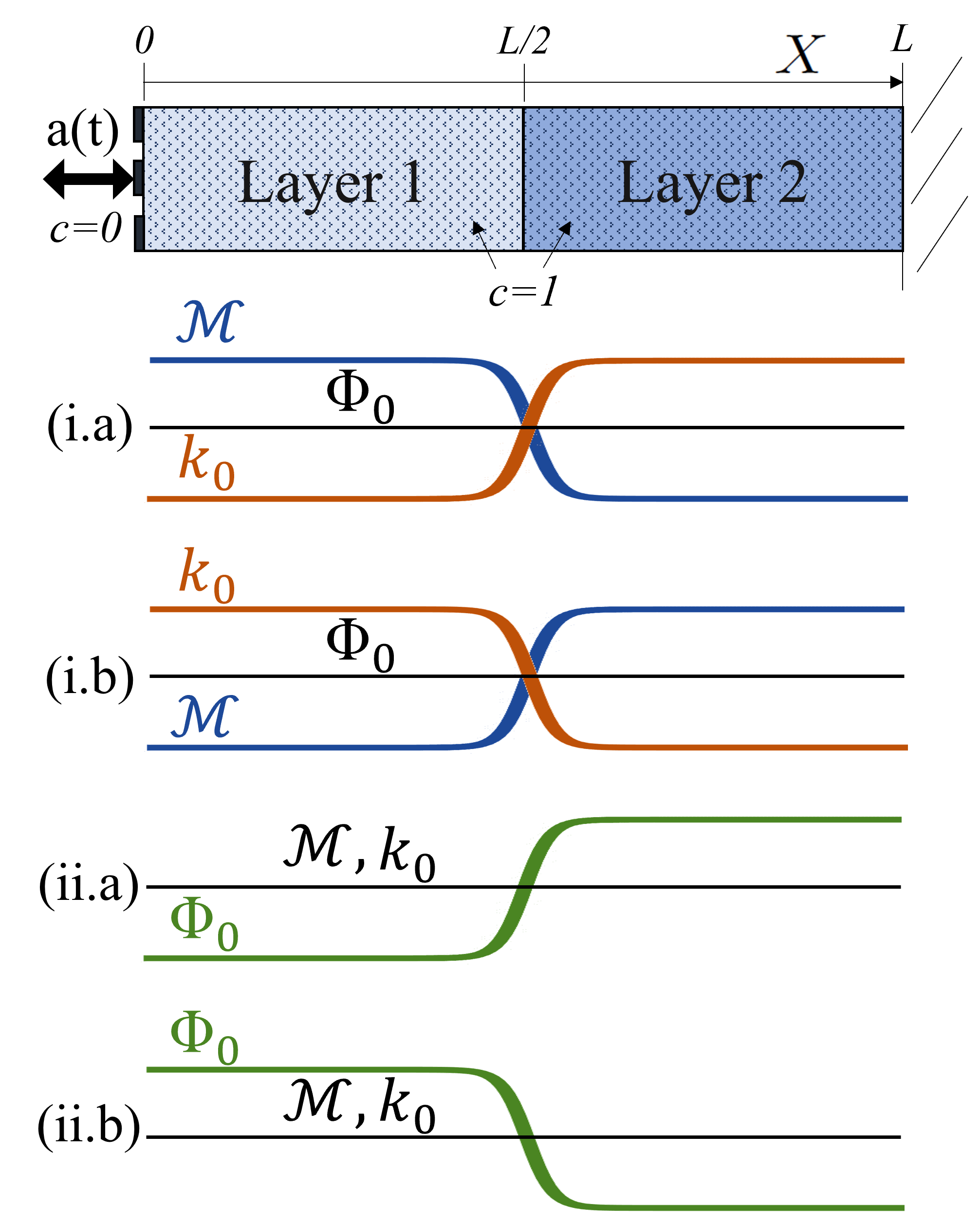}
    \caption{Schematic representation of a layered medium subject to periodic, displacement-driven load, through a permeable piston (left boundary), whereas the right boundary is fixed and impermeable. The concentration of solutes is $C=1$ in the domain ($X\in[0,L]$), and $C=0$ outside. Below, we highlight the four bilayer cases studied (properties not to scale).  }
    \label{fig:scheme}
\end{figure}

In homogeneous media subject to pulsatile loads, for fixed loading amplitude $A$ and period $T$, the material response is driven by the poroelastic timescale:
\begin{equation}\label{Tpe}
T_{\mathrm{pe}} = \frac{\mu L^2}{k_0 \mathcal{M}},
\end{equation}
defined by $\mu$, the fluid viscosity, $L$ the domain length, and $k_0$ and $\mathcal{M}$, the material resting permeability and \textit{p}-wave modulus, respectively. Importantly, the initial porosity $\Phi_0$ does not impact $T_{\mathrm{pe}}$.
Varying $T_{\mathrm{pe}}$ significantly modifies the spatial dependence of the material response, with two asymptotic regimes: a slow-loading regime ($T\gg T_{\mathrm{pe}}$), with no strain/flow localisation, and a fast-loading regime ($T\gg T_{\mathrm{pe}}$) with emergence of localisation~\citep{Fiori2023}. To isolate the role of macroscopic heterogeneities, we filter out these $T_{\mathrm{pe}}$-dependent effects, by varying the individual layer properties under two constraints: (i) each layer is assigned the same $T_{\mathrm{pe}}$ (through a constant product $k_0 \mathcal{M}$) to prevent contrasting mechanical regimes within the material halves; and (ii) a constant effective macroscopic $T_{\mathrm{pe}}$ is maintained across the entire material, computed via the harmonic average of $k_0$ and $\mathcal{M}$ across the layers.
We consider four scenarios (Fig.~\ref{fig:scheme}): we either co-vary $\mathcal{M}$ and $k_0$ at constant $\Phi_0=0.5$, or vary $\Phi_0$ at fixed $\mathcal{M}=k_0=1$. 
Table~\ref{table:cases} lists the resulting layer-specific parameters.
\begin{table}
\centering
\begin{tabular}{l|c|c}
\textbf{Case}              & Layer 1   ($\mathcal{M}, k_0, \Phi_0$)     & Layer 2 ($\mathcal{M}, k_0, \Phi_0$)     \\ 
\hline
Homogeneous  & $1, 1, 0.5$ & $1, 1, 0.5$ \\
(i.a)              & $3, 0.6,0.5$  & $0.6, 3, 0.5$ \\
(i.b)        & $0.6, 3, 0.5$ & $3, 0.6, 0.5$ \\
(ii.a) & $1, 1, 0.25$ & $1, 1, 0.75$ \\
(ii.b) & $1, 1, 0.75$ & $1, 1, 0.25$ \\ 
\end{tabular}
\caption{Summary of the dimensionless parameters explored in each case, as schematised in Fig.~\ref{fig:scheme}.  }
\label{table:cases}
\end{table}
We set $\mathcal{M}$, $k_0$, $\Phi_0$ individually, as the interdependencies between these parameters are non-trivial given the complex microstructure of soft biological tissues~\cite{HRABETOVA2004, Nicholson23}.
For instance, an increased $\Phi_0$ does not intrinsically imply a higher $k_0$, if the restrictive dimensions of the pore throats remain unchanged~\cite{Nelson1994, Katz86}. Notably, while we treat the initial parameters as independent, the deformation-dependent permeability is modelled by a Kozeny-Carman relation (see equation~\eqref{kc-permeability} below), where the current permeability $k(X,\phi_f)$ evolves as a function of the current true fluid fraction $\phi_f$ relative to  $\Phi_0$, with $k_0\equiv{}k(X,\phi_{f,0})$.

The domain consists of a bilayer medium of total length $L$, partitioned into two sub-regions of equal thickness: $0 \leq X \leq L/2$ and $L/2 \leq X \leq L$. 
Note that, while the length ratio undoubtedly influences the overall poromechanical behaviour during simultaneous parameter shifts, the current study is designed to isolate the specific effects of layered heterogeneities. This simplified configuration serves as an optimal baseline for this objective. 

The medium is subject to a periodic displacement of period $T$ at a permeable left boundary, while the right boundary is fixed and impermeable. The initial symmetry in the layers does not imply equal thickness during loading, as the spatial variations in poroelastic properties induces non-uniform deformations in the two layers.
To avoid interfacial discontinuities, we smoothly transition each initial property $f(X, 0)$ (representing $\mathcal{M}$, $k_0$, or $\Phi_0$) between its respective values $f_1$ and $f_2$ using a hyperbolic tangent function $f(X, 0) = f_1 + \frac{f_2 - f_1}{2} \left[ 1 + \tanh \left( \kappa \left( X - \frac{L}{2} \right) \right) \right]$, where the steepness coefficient is set to $\kappa=100$ to ensure a rapid yet differentiable transition across the interface at $X = L/2$ (see Fig.~\ref{fig:scheme}).
Given such spatial variations, we reformulate the model from \citep{Macminn2016} in Lagrangian coordinates, following~\cite{Fiori_thesis, Godard25}.

\section{Model}
In this section, we summarise the equations defining the model for poromechanics and solute transport in Lagrangian coordinates. 
\subsection{Kinematics}
The Lagrangian solid displacement field is $\mathbf{U_{s}}(\mathbf{X},t)=\mathbf{x}(\mathbf{X},t)-\mathbf{X}$,
with $\mathbf{x}(\mathbf{X},t)$ the current position of the material point with reference position $\mathbf{X}$. 
Our reference configuration is the relaxed configuration, where $\mathbf{x}(\mathbf{X},0)=\mathbf{X}$, so that $\mathbf{U_{s}}(\mathbf{X},0)=0$.
Given the Jacobian determinant $J= 1+\partial{U_s}/\partial{X}$, for incompressible constituents and uniform initial porosity $\Phi_{f,0}$, the local change in volume relates to the change in porosity as
\begin{equation}\label{porosity-def-lagr}
    J(X,t) = 1-\Phi_{f,0}+\Phi_f \quad\to\quad \frac{\partial{U_s}}{\partial{X}} =\Phi_f-\Phi_{f,0},
\end{equation}
where the nominal porosity $\Phi_f=J\phi_f$ measures the current fluid volume per unit reference total volume and is expressed as function of the true porosity $\phi$.

\subsection{Darcy's law}
Mass conservation for the fluid phase reads:
\begin{equation}\label{continuity-lagr}
    \frac{\partial{\Phi_f}}{\partial{t}} +\frac{\partial}{\partial{X}}(W_f) = 0,
\end{equation}
where $W_f$ is the nominal flux of fluid through the solid skeleton. 
By definition, $W_f$ is given by Darcy's law, which reads 
\begin{equation}
W_ f = - \frac{1}{J}\frac{k(X,\phi_f)}{\mu}\frac{\partial p}{\partial X},
\label{darcylaw-lagr}
\end{equation}
where $p$ is the pore pressure.
For the permeability law, we choose a normalised Kozeny-Carman relation of the form
\begin{equation}\label{kc-permeability}
    k(\phi_f) = k_0\,\frac{(1-\phi_{f,0})^2}{\phi_{f,0}^3}\, \frac{\phi_f^3}{(1-\phi_f)^2},
\end{equation}
where $k_0\equiv{}k(X,\phi_{f,0})$ is the permeability of the initial state, as mentioned above.

\subsection{Mechanical equilibrium}
In 1D, mechanical equilibrium reads: 
\begin{equation}
    \frac{\partial s}{\partial X}=0,
\label{s}
\end{equation}
where $s$ is the nominal total stress. The nominal effective stress $s'$ is then:
\begin{equation}
     s'= s+p.
\label{s'}
\end{equation}
We take the elasticity law to be Hencky elasticity~\cite{Hencky33}: 
\begin{equation}\label{sigma-hencky-lagr}
    s^\prime= \mathcal{M}\frac{\ln(J)}{J}.
\end{equation}

\subsection{Fluid flow}
Combining mass conservation, Darcy's law, and mechanical equilibrium, the final nonlinear flow equations read: 
\begin{equation}
    \frac{\partial{\Phi_f}}{\partial{t}} -\frac{\partial}{\partial{X}}\bigg[\frac{1}{J} \frac{k(X,\phi_f)}{\mu}\frac{\partial{p}}{\partial{X}}\bigg]=0 \quad\mathrm{and}\quad \frac{\partial{s'}}{\partial{X}}=\frac{\partial{p}}{\partial{X}}.
\label{conservation-q-lagr}
\end{equation}
Note that the fluid and solid velocities ($v_f$ and $v_s$ respectively) are the same in Eulerian and Lagrangian space. 
Since 
\begin{equation}\label{wf-velocities}
 W_f= \phi_f (v_f - v_s),
\end{equation}
it is easy to derive that 
\begin{equation}\label{vs-vf}
    v_s=-W_f\quad\mathrm{and}\quad v_f= - \left( \frac{1-\phi_f}{\phi_f}\right) v_s.
\end{equation}

With appropriate initial and boundary conditions, the above formulations close the Lagrangian model.

\subsection{Solute transport}
The model for solute transport is analogous to the one presented in \cite{Fiori2025}. In a Lagrangian frame, conservation of mass for the solute reads: 
\begin{equation}
\begin{split}
 \frac{\partial (\Phi_f C)}{\partial t}  +\frac{\partial}{\partial X}\bigg[{ C W_f }- \frac{1}{J}\phi_f \mathcal{D}   \frac{\partial C}{\partial X}\bigg]=0, \\
 \textrm{with }\quad \mathcal{D}   =\mathcal{D}_m+\mathcal{D}_h = \mathcal{D}_m+ \alpha \bigg| \frac{W_f}{\phi_f} \bigg|, 
 \end{split}
\label{conservation-c-lagr}
\end{equation}
where $C$ represents the intrinsic solute concentration and the diffusivity $\mathcal{D}$ is written in terms of the coefficients for molecular diffusion $\mathcal{D}_m$ and for hydrodynamic dispersion $\mathcal{D}_h$ using the dispersivity $\alpha$. The total solute flux is 
\begin{equation}
W_c=C W_f - \frac{1}{J}\phi_f \mathcal{D}   \frac{\partial C}{\partial X}.  
\end{equation}

\subsection{Boundary conditions}
We define the left and right boundaries of the solid to be at $X=0$ and $X=L$, respectively, with initial conditions for the mechanics
$\Phi_f(X,0)= \phi_{f,0}$  and $U_s(X,0)=W_f(X,0)=0 $. For the solute, in analogy with brain clearance, we initially impose $C(X,0)=1$ throughout the domain, meaning that solutes are initially present inside the medium and can escape through the left boundary into a solute-free fluid reservoir.
At the left boundary, we apply a periodic displacement $a(t)= \frac{A}{2} \left[1-\cos\left(\frac{2\pi t}{T}\right)\right]$, with boundary conditions:
\begin{equation}\label{bc-left}
        v_s(X=0,t) = \dot a(t),\quad  p(X=0,t)=0      
\end{equation}
for the mechanics, and
\begin{equation}
 C(0,t)=0,       
\end{equation}
for solute transport. 
The right boundary is impermeable and fixed in place, such that
\begin{equation}\label{bc-right}
    U_s(L,t)=W_f(L,t)=0,
\end{equation}
and
\begin{equation}\label{bc-right-st}   \frac{\partial C}{\partial X}\Bigg|_{X=L}=0.
\end{equation}
We fix the values of the loading amplitude and period to typical mean values for biological media, with $A/L=0.05$ and $T/T_{\mathrm{pe}}=1$ (see \cite{Fiori2023}); analogously, we take a typical dimensional diffusion coefficient $\mathcal{D}_m=10^{-10}$m$^2$/s \cite{Fiori2025}, and we vary the value of $\alpha$.

\subsection{Scaling}
We report here the model in dimensionless form. 
The evolution equations are nondimensionalised according to:
\begin{equation}
\begin{split}
    \tilde{X}=\frac{X}{L},\;
    \tilde{U}_s=\frac{U_s}{L},\;
    \tilde{t}=\frac{t}{T_{\mathrm pe}},\;
    \tilde{s}^\prime=\frac{s'}{\mathcal{M}},\;
    \tilde{p}=\frac{p}{\mathcal{M}},\; \\\tilde{k}=\frac{k(\phi)}{k_0},\; 
    \tilde{W}_f=\frac{W_f}{L/T_{\mathrm pe}}, 
\end{split}
\end{equation}
for the poromechanics part, and
\begin{equation}
    \tilde{C}=\frac{C}{C_0},  \; 
   \tilde{\alpha}=\frac{\alpha}{L},\;
    \tilde{\mathcal{D}}=\frac{\mathcal{D}}{\mathcal{D}_m}=1+\tilde{\alpha} \mathrm{Pe}\bigg | \frac{\tilde{W}_f}{\phi_f} \bigg |
\end{equation}
for the solute transport part, where $\mathrm{Pe}=(L^2/\mathrm{T}_{pe})/\mathcal{D}_m$ is the P\'eclet number.
The full problem can then be rewritten in dimensionless form as
\begin{equation}
     \frac{\partial{\Phi_f}}{\partial{\tilde{t}}} -\frac{\partial}{\partial{\tilde{X}}}\left(\tilde{W}_f\right)=0
\label{conservation-q-scaled-lagr}
\end{equation}
and 
\begin{equation}
 \frac{\partial (\Phi_f \tilde{C})}{\partial \tilde{t}}  +\frac{\partial}{\partial \tilde{X}}\bigg[{\tilde{C} \tilde{W}_f }- \frac{1}{J} \phi_f \mathrm{Pe}^{-1}\tilde{\mathcal{D}}  \frac{\partial \tilde{C}}{\partial \tilde{X}}\bigg]=0,
\label{conservation-c-scaled-lagr}
\end{equation}
with initial conditions
\begin{equation}
\begin{split}
   \tilde{a}(0)=0, \;
   \Phi_f(\tilde{X},0)= \phi_{f,0}, \; 
    \tilde{W}_f(\tilde{X},0)=\tilde{v}_s(\tilde{X},0)=0, \;\\
    \tilde{C}(\tilde{X},0)=1.
    \end{split}
\end{equation}
The scaled left boundary conditions read:
\begin{equation} \label{leftB-cc}
      \begin{split}
\tilde{U}_s(0,\tilde{t})=\tilde{a}(\tilde{t})= \frac{\tilde{A}}{2} \Bigg[1-\cos\left(\frac{2\pi\tilde{t}}{\tilde{T}}\right)\Bigg]
    \,,\\\,\,\tilde{W}_f(0,\tilde{t})=-\tilde{\dot{a}}(t)\,
    \quad\mathrm{and} \quad \tilde{C}(0,\tilde{t})=0,
    \end{split}
\end{equation}
where $\tilde{A}=A/L$ and $\tilde{T}=T/T_{\mathrm{pe}}$. 
The right boundary conditions are
\begin{equation}
    \tilde{U}_s(1,\tilde{t}) =\tilde{W}_f(1,\tilde{t})=0 \quad\mathrm{and}\quad
    \frac{\partial  \tilde{C}}{\partial \tilde{x}}\Bigg|_{\tilde{x}=1} =0.
\end{equation}
In the text, we refer to dimensionless values dropping the tildas for convenience everywhere except for the example of brain GM and WM, where dimensional parameters are essential to define the problem.

We solve our model numerically in MATLAB, using compact finite differences in space, with implicit Runge-Kutta integration in time, which we achieve via
MATLAB’s built-in solver ODE15s~\cite{shampine-siamjscicomput-1997}. The full description of the numerical method and the converge analsysi can be found in~\cite{Fiori2025}.

\section{Results}
\subsection{Impact of layers on poromechanics}
\begin{figure*}[]
  \centering
  \includegraphics[width=0.65\textwidth, trim=3.5cm 0.cm 3.5cm 0.cm]{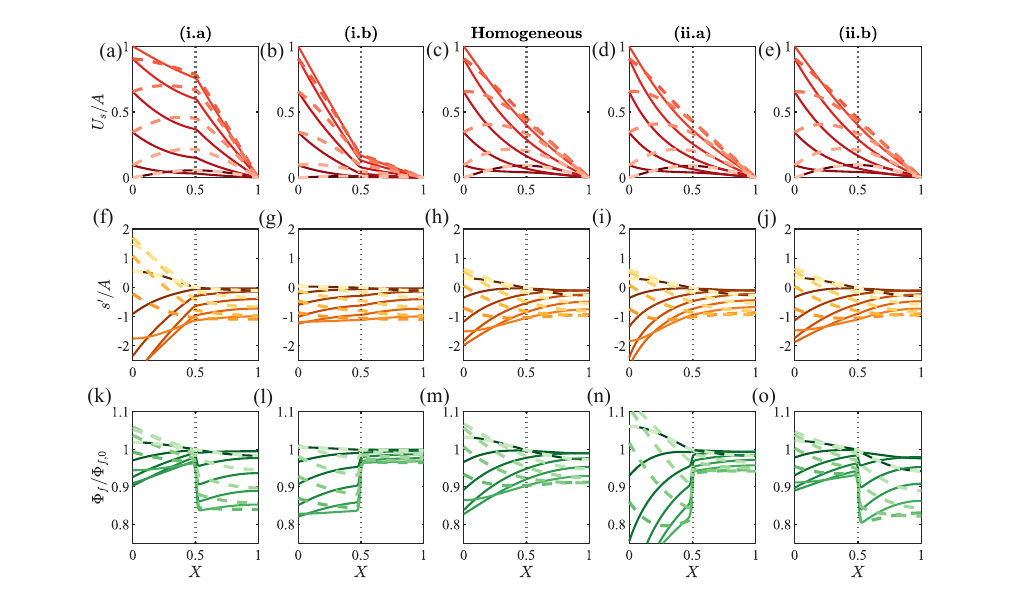}
  \caption{Space distribution of the normalised displacement $U_s$, solid stress $s^\prime$, and local variation in porosity $\Phi_f/\Phi_0$ at ten equally spaced time-steps in a loading cycle (dark to light), during loading vs unloading (solid vs dashed lines), for the five cases of bilayer media summarised in table~\ref{table:cases}.}
  \label{fig:mechanics}
\end{figure*}
\begin{figure*}[]
  \centering
  \includegraphics[width=0.65\textwidth, trim=3.5cm 0.cm 3.5cm 0.cm]{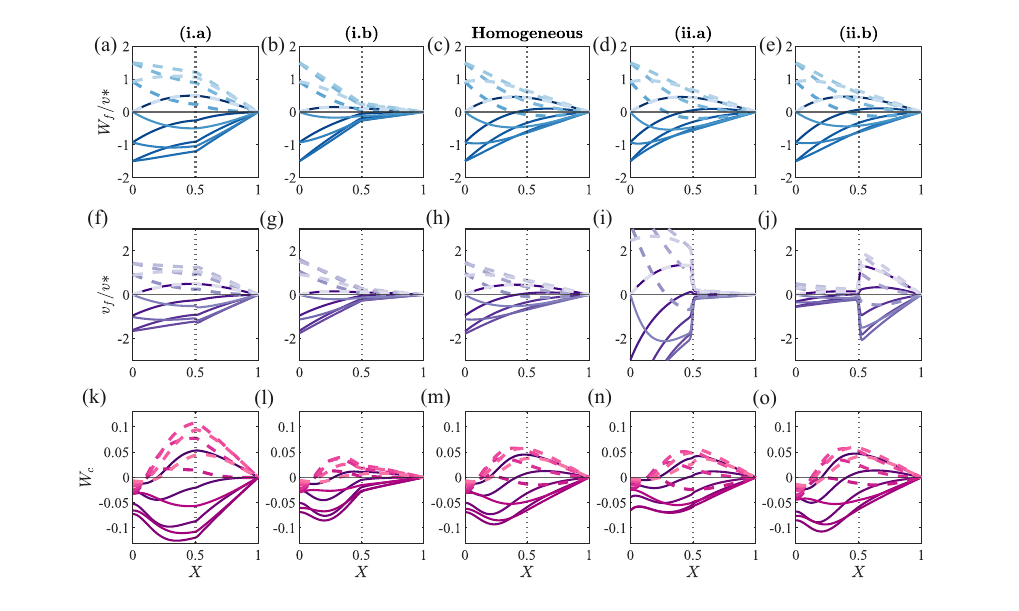}
  \caption{Space distribution of the normalised fluid flux $W_f$, fluid velocity $v_f$, and solute flux $W_c$ at ten equally spaced time-steps in a loading cycle (dark to light), during loading vs unloading (solid vs dashed lines), for the five cases of bilayer media summarised in table~\ref{table:cases}. $W_c$ is computed at the second cycle, with $\alpha=0.05$.}
\label{fig:flow}
\end{figure*}

We evaluate the impact of macroscopic heterogeneity on poromechanics and solute transport by comparing five configurations: a homogeneous control and the four bilayer cases described in table~\ref{table:cases}.  Fig.~\ref{fig:mechanics} displays the corresponding displacement $U_s$, solid stress $s^\prime$, and local change in porosity $\Phi_f/\Phi_0$, and Fig.~\ref{fig:flow} the fluid flux $W_f$, fluid velocity $v_f$ (both normalised by $v^\star=2A/T$), and solute flux $W_c$. 
The presence of layers fundamentally alters the spatial distribution of mechanical and transport quantities. Notably, the system response is not solely a function of its average properties, but is critically dependent on the spatial arrangement of the layers relative to the pulsatile boundary. 
In details, in case (i.a), the first layer’s high impedance ($\uparrow \mathcal{M}, \downarrow k_0$) resists local deformation, generating high solid stresses that effectively transmit strain into the second layer (Fig.~\ref{fig:mechanics}(a-f-k)). This propagation effect amplifies fluid and solute fluxes deep into the medium (Fig.~\ref{fig:flow}(a-f-k)). Conversely, in case (i.b), the high deformability of the superficial layer ($\downarrow \mathcal{M}, \uparrow k_0$) localises strains and fluid flow near the piston, significantly attenuating distal solute transport (Fig. \ref{fig:mechanics}(b-g-l) and \ref{fig:flow}(b-g-l)). 
Variations in $\Phi_0$ (series ii) do not impact displacement and fluid flux spatial propagation. However, a lower proximal porosity (ii.a) enhances localization (Fig.~\ref{fig:mechanics}(i-n)), characterised by higher velocities and solid stresses in the proximal layer, but diminished total solute transport (Fig.~\ref{fig:flow}(i-n)). In contrast, increasing proximal porosity (ii.b) mitigates these localizations compared to the homogeneous case, thereby enhancing overall solute clearance (Fig. \ref{fig:mechanics}(j-o) and \ref{fig:flow}(j-o)).  

\subsection{Impact of layers on solute transport}
\begin{figure}[h!]
     \centering
     \includegraphics[width=0.7\linewidth, trim=0.25cm 0.cm 0.25cm 0.cm]{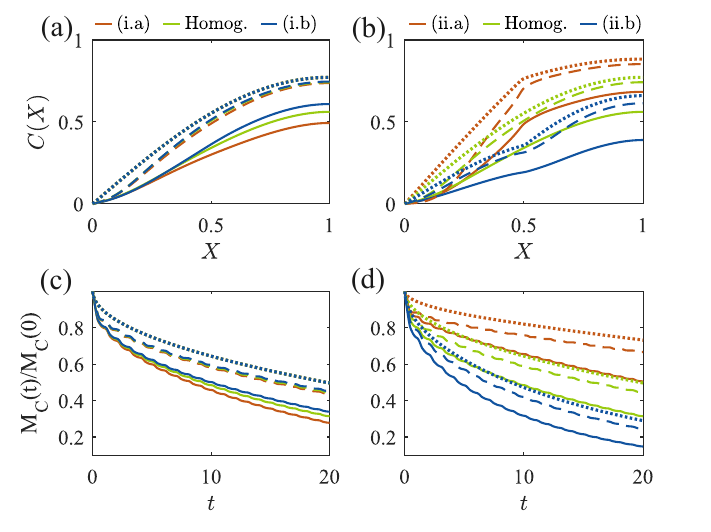}
     \caption{
     Solute transport dynamics for the five cases of bilayer media summarised in table~\ref{table:cases}. Final concentration profiles (a)-(b) and normalised total solute mass over time (c)-(d) for mechanical property variations (a)-(c) and porosity variations (b)-(d), compared against the homogeneous baseline.  Dotted lines show pure diffusion ($\alpha=0, A=0$). Dashed and solid lines represent advection-diffusion at $\alpha=0$ and advection-diffusion-dispersion at $\alpha=0.05$, respectively.}
     \label{fig:concentrations}
 \end{figure}

In Fig.~\ref{fig:concentrations}, we quantify the impact of heterogeneities on solute transport. Considering 20 loading cycles, we show the final concentration profile across the domain (Fig.~\ref{fig:concentrations}(a-b)) and the total solute amount over time (Fig.~\ref{fig:concentrations}(c-d)). For each case, we consider diffusion only, advection-diffusion (no dispersion, $\alpha=0$), and advection-diffusion-dispersion ($\alpha=0.05$). Cases (i.a) and (ii.b) show a lower final solute concentration compared to the homogeneous case, \textit{i.e.}  enhanced clearance. On the other hand, case (i.b) and (ii.a) display higher solute concentration, highlighting an overall reduced solute transport.
In series (i), such transport dynamics are intuitive: proximal resistance to deformation reduces localisation in (i.a) vs (i.b), enhancing solute transport, with no impact in pure diffusion, as expected. 
Conversely, series (ii) reveals a more complex interplay of poromechanical and diffusive effects. Even in the diffusion-only limit, the higher porosity near the exit in (ii.b) accelerates clearance by increasing the concentration gradient. Under periodic loading, moreover, series (ii) exhibits counter-intuitive behaviour: proximal resistance (low $\Phi_0$ near the piston) increases localization in fluid volume changes, further hindering transport in (ii.a).
Consequently, transport physics is dictated by the resulting spatial distribution of fluid volume changes rather than proximal deformability. Configurations (i.a) and (ii.b) are the most favourable for clearance; by minimizing localization and shifting deformation toward the distal end, these setups maximize fluid and solute fluxes deep into the tissue.

\section{Biological bilayer system: brain grey and white matter}
We next implement this bilayer poromechanical analysis in a case study of amyloid-$\beta$ transport within the grey matter (GM) and white matter (WM) system in the brain.
In this part, $T_{\mathrm{pe}}$ is not constant anymore and depends on the tissue parameters. On average, the thickness of the GM is $2.5$mm~\cite{Fischl2000}, and its surface is subject to pulsations from pial arteries. Such pulsations have the largest amplitude at the frequency of vasomotion~\cite{VanVeluw2020, BROGGINI2024}, and are likely to propagate from the GM surface to the WM surface, since parenchymal volumetric strain linked to cardiac pulsations is documented in those regions~\cite{Adams2019, SLOOTS2021, Sloots2022}. Hence, we define our domain as a bilayer of total length $5$mm, equally partitioned between GM and WM. We take $T=10s$ (dimensional value, corresponding to the typical vasomotion frequency of $0.1Hz$) and $A=15$$\mu$m, since vasomotion leads to strains in the order of $15 -20\%$ of the vessel diameter (with typical diameter values of $200-250$$\mu$m~\cite{Bollmann22}). Reference properties of healthy GM and WM are taken from literature (post-mortem experiments) and are summarised in table~\ref{table:props_GMWM}. The ratio $\mathcal{M}_{GM}$ to $\mathcal{M}_{WM}$ is assumed to be the same as reported for the Young modulus~\cite{BUDDAY2015}.
For the pathological changes, we focus on the impact of amyloid-$\beta$ accumulations, characterising the GM only; hence, we do not consider any changes in the WM properties. We assume that pathological alterations of the GM lead to: (i) a decrease in tissue stiffness, due to neuronal death~\cite{Murphy2011, MURPHY2016, Hiscox2019}; (ii) a decrease in permeability, due to the progressive clogging of the interstitial fluid space~\cite{YUE2019, Iorio24}, usually reported as increased tortuosity~\cite{HRABETOVA2004, Sykova2005}; and (iii) a progressive decrease in porosity, due to reductions in functional pore space connectivity. 
Assumption (iii) is related to the experimental observations that amyloid plaques create ``dead-space" microdomains that trap extracellular water~\cite{HRABETOVA2004} and that inflammated astrocytes swell~\cite{Nicholson1998}, reducing the effective GM fluid fraction actively participating in clearance. Together with the increased cortical free water observed via MRI in Alzheimer's disease \cite{Hall2026}, these observations imply that while more extracellular water is present, a larger portion remains stagnant. 
In details, for the GM pathological states P1 and P2,  we assume $\mathcal{M}$ and $k_0$ to be respectively half and one quarter of their healthy value; $\Phi_0$ is reduced of one third and half of the reference healthy value (as summarised in table~\ref{table:props_GMWM}).

For the solute transport, we take the dimensional amyloid-$\beta$ diffusion coefficient to be $\mathcal{D}_m=6\times10^{-11}$m$^2$/s~\cite{Waters2011}.
Values for the dispersivity $\alpha$ usually scale with the size of the sources of heterogeneity in the medium. In a homogeneous gel, for example, $\alpha$ is set up by the size of the gel pores~\cite{Noh2024}. 
At the meso-scale considered here, the brain tissue is highly heterogeneous due to the presence of several different structures (such as, blood vessels of different sizes, perivascular spaces, cells). Hence, we choose dimensional dispersivity values of $\alpha = 50\text{ and }250\ \mu\text{m}$, ranging from the typical length-scale of a single capillary vessel ($50\ \mu\text{m}$)
to the characteristic spacing between
large penetrating vessels and their perivascular space ($100\text{ to }300\ \mu\text{m}$), to account for the macroscale hydrodynamic spreading \cite{DUVERNOY1981, Lorthois2011}.
\begin{table*}
\centering
\begin{tabular}{l|c|c|c|c}
\textbf{Tissue}              & $\mathcal{M}$   [kPa]       & $k_0$ [m$^2$]    & $\Phi_0$ & $T_{\mathrm{pe}} $ [s] (single layer) \\ 
\hline
WM - Healthy                  & $250$ \cite{Franceschini2006}                         & $5\times10^{-15}$ & $0.2$~\cite{Lehmenkuhler1993}  & 5    \\
GM - Healthy        & $0.7\times\mathcal{M}_{WM}$\cite{BUDDAY2015}$=175$ & $4\times10^{-15}$\cite{Greiner24} & $0.22$~\cite{Lehmenkuhler1993}    & 9    \\
GM - P1 & $131.25$                        & $2\times10^{-15}$  & $0.15$ &24     \\
GM - P2 & $87.5$                      & $1\times10^{-15}$ & $0.11$   &70 
\end{tabular}
\caption{Values considered in the model for WM, GM, and the GM pathological alterations P1 and P2 due amyloid-$\beta$ plaques. Note that in the Pathological GM cases, alterations of the parameters are speculative as they are challenging to find in literature.   }
\label{table:props_GMWM}
\end{table*}

\begin{figure}[H]
    \centering
    \includegraphics[width=0.7\linewidth, trim=0.25cm 0.cm 0.25cm 0.cm]{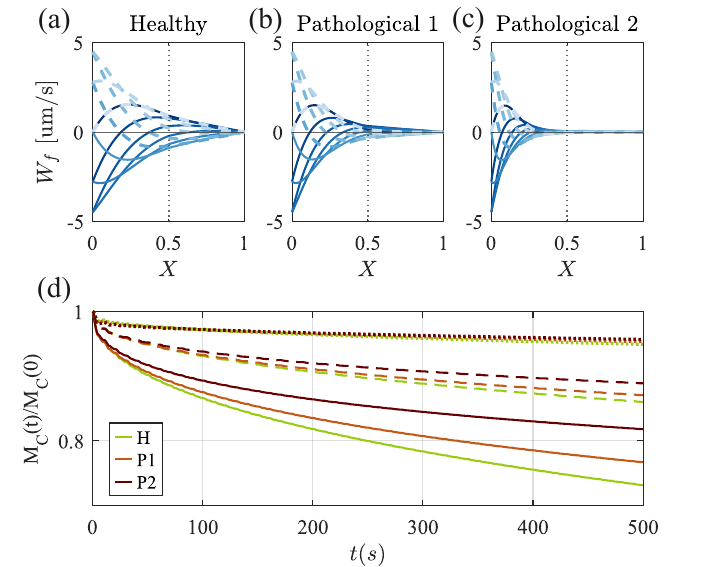}
    \caption{Poromechanics and solute transport of the GM-WM system, with healthy vs pathological properties (see table~\ref{table:props_GMWM}). Space distribution of $W_f$ (a)-(b)-(c) (dimensional), at equally spaced time-steps in a loading cycle (dark to light), during loading vs unloading (solid vs dashed lines). Temporal evolution of the normalised solute mass (d) for varying $\alpha$, with $\alpha=0$ (dotted lines), $\alpha=0.01$ (dashed lines) and $\alpha=0.05$ (solid lines). }
    \label{fig:GM-WM}
\end{figure}

Fig.~\ref{fig:GM-WM}(a-b-c) summarises the impact of the pathological changes on the mechano-fluid properties of the GM-WM system. 
The progressive decrease in stiffness and permeability in the GM layer leads to progressively stronger localisation of the fluid flux at the surface: in the healthy condition, $W_f$ spreads through the WM, while in the pathological cases it is almost entirely suppressed in the WM and strongly reduced in the distal portions of the GM.
Such changes in fluid-flow distribution impact on solute transport dynamics (Fig. \ref{fig:GM-WM}(d)), noticeably impaired in both pathological cases. 
Given the narrow physio-pathological range assumed for $\Phi_0$ (see table~\ref{table:props_GMWM}), its impact is minor and only marginally observable in the absence of dispersion ($\alpha=0$). We consider this scenario to be physiologically improbable, however, as the profound microstructural heterogeneity of brain tissue \cite{HRABETOVA2004, Nicholson23} is expected to induce significant hydrodynamic dispersion~\cite{Fiori2025}.
\section{Conclusions}
In this work, we have explored the pulsatile poromechanics of layered soft porous media, and its implications on solute transport. Our results reveal a clear dependence on the spatial sequencing of the layers relative to the loading boundary. This asymmetry arises from the phase lag between tissue deformation and pore-pressure relaxation, suggesting that the order of tissue heterogeneities is as critical as the properties themselves. Such interactions among layers lead to complex distributions of fluid fractions, stresses, and fluid fluxes, which are not achievable through classic changes in mechanical regimes (governed by the poroelastic timescale~\cite{Fiori2023}) in a homogeneous medium. 

Furthermore, our findings demonstrate that solute transport dynamics is strongly influenced by bilayer architectures. This insight is of fundamental significance for modelling transport phenomena in biological media.
From a biophysical aspect, these results suggest that the division in layers could hence be a strategy for each tissue to reach an efficient response to fixed oscillating boundary conditions, such as blood vessel pulsations.
Moreover, these results could guide the design of ``smart'' tissue engineering scaffolds: tuning their local stress/flow while controlling local transport of nutrients and/or waste substances could dramatically improve cell survival and proliferation. 

In the final section of this work, we extend the developed framework to a physiologically relevant case: the grey and white matter bilayer of cerebral tissue. In this case, the empirical parameters sourced from literature naturally deviate from the uniform poroelastic timescales assumed in our theoretical analysis, but the model presented here is still fully valid. 
Although simplified, our 1D approach offers a fundamentally new perspective on brain clearance by isolating the interfacial physics of bilayer poroelasticity. We show that altering the mechanical balance between layers --- even in this idealized limit --- is sufficient to reduce clearance efficiency. This suggests that the brain’s architecture is inherently sensitive to pathological shifts in material properties: a pathological GM is likely to further hinder clearance because of the altered poromechanical equilibrium, thus generating a self-reinforcing ``vicious circle" for waste accumulation~\cite{Greenberg2020}. Importantly, while qualitatively based on published literature, the pathological changes assumed here are speculative and would require experimental verification to advance our understanding of brain poromechanics. Specifically, local \textit{in vivo} measurements of permeability, stiffness, and porosity --- across both healthy tissues and different stages of pathological progression --- are essential to rigorously characterize the biophysical coupling between poromechanics and solute transport. In this regard, more precise assessments of the specific transport mechanisms happening \textit{in vivo}, such as the dynamic interplay between deformation-dependent diffusion and dispersion, remain highly necessary to validate upscaled continuum models.

In future work, we plan to extend the bilayer parameter space exploration, and to focus on the upscaling of solute transport to retrieve effective transport estimations for each layered case. Furthermore, we believe it would be interesting to refine our model of the cerebral grey matter–white matter system by incorporating dual-porosity poroelasticity, a framework particularly suited for brain tissue~\cite{Sciume2021}. 

\vspace{\baselineskip}
\textit{Acknowledgements}:
This work was supported by a grant from the
Leducq Foundation and the Leducq Foundation for Cardiovascular
Research (23CVD03). We thank the members of the  Leducq Network for the stimulating environment for exchanges and interdisciplinary discussions.
We are grateful to C. W. MacMinn for the inspiring discussions and for the assistance with the formulation of the model in Lagrangian coordinates. We also thank S. Waters, D. Moulton, and Z. Godard for insightful discussions.

 \bibliographystyle{apsrev4-2}
\bibliography{references.bib}       
\clearpage


\end{document}